\newcommand{\pot}{V_{q\bar q}}
\newcommand{\LMSB}{\Lambda_{\overline{MS}}}
\newcommand{\LR}{\Lambda_R}

\documentstyle[twoside,fleqn,espcrc2]{article}

\title{Pure-gauge SU(2) on large lattices}

\author{UKQCD collaboration, presented by P.W. Stephenson\address{%
	Department of Applied Mathematics and Theoretical Physics,
	Chadwick Tower, \\
	University of Liverpool,
	P.O. Box 147, Liverpool L69 3BX, United Kingdom \\
	Email: {\tt pws@s-a.amtp.liv.ac.uk}}}

\begin{document}

\begin{abstract}%
We have calculated the $q\bar q$ potential over a wide range of lattice
separations on a $48^3\cdot56$ lattice at $\beta = 2.85$.  We are able to
investigate both long-range and (by correcting for the effects of lack
of rotational invariance) short-range potentials.  From the former we
estimate the string tension and from the latter we are able to
investigate asymptotic scaling.  Fitting to this enables us to give an
estimate for the $\Lambda$-parameter of pure-gauge SU(2).
\end{abstract}

\maketitle

\section{Introduction}
We discuss recent results for the potential between (hypothetical)
static quarks calculated on the 64-node Meiko Computing Surface at
Edinburgh University.  Our lattice size, $48^3\cdot56$, is to our
knowledge the largest so far used in a lattice gauge theory simulation.
Thus we are in a good position to examine the approach to the continuum
limit and comment on the remaining effects of finite lattice spacing
$a$.  A fuller write-up is available as a preprint~\cite{us}, so after
briefly mentioning our data and techniques we proceed directly to a
discussion of the results.

\section{Monte Carlo data}
We have chosen $\beta=2.85$ based on previous experience to make our
lattice physically as large as necessary.  It is known that the
potential is less sensitive to finite-volume effects than, for example,
the glueball spectrum but we do not wish to rely on this.  We also have
results for glueballs which will not be discussed here.  We have
performed a total of over $10^{12}$ link updates on this lattice.

Our technique for extracting the potential $\pot$ is standard. The
effective potential in lattice units is extracted from $R\times T$
Wilson loops as $V_{\rm eff}(R,T)a = \ln(W(R,T)/W(R,T-1))$; as $T$
becomes large, $V_{\rm eff}(R,T)$ tends to the ground state potential
$\pot(R)$ which is what we need.  To improve our overlap onto this state
we have used two popular `fuzzing' schemes~\cite{teperblock,apefuzz};
the `wave functions' generated have been used as the basis for a
variational analysis.  The resulting overlap approaches unity in many
cases.

We are faced as always with the problem of deciding how to extract
$\pot(R)$ from $V_{\rm eff}(R,T)$.  We find that the results at $T=5$ agree
within statistical errors with those at $T=4$ and use these as our main
results.  We have also performed an extrapolation to $T=\infty$ by
assuming that the the signal is contaminated only by one excited state.
This assumption gives a lower bound on the potential.  In this case we
are less in control of statistical errors; furthermore, we expect this
procedure to be less effective for larger $R$ where the first excited
state is closer to the ground state.  Thus we use the extrapolated
potentials mainly as a way of gauging our systematic errors.

\section{Fits to data}

We have performed two separate fits for large and for small separations
$R$.  In fact, we have used different data analysis techniques in the
two cases: the large-separation analysis required larger statistics but we
reduced the total amount of data produced by blocking the lattice by a
factor of two in each spatial direction.  In the other case we needed
the full lattice volume but lower statistics for accurate results.

\subsection{Large $R$}
Here our major object is to investigate the string tension $K$.  We fit to
the form
\begin{equation}
	V(R) = C - {E\over R} + KR
\label{eq:bigfit}
\end{equation}
for $R\ge6a$.  In fact, the Coulomb term is modified to include
corrections (from first order perturbation theory) for lack of
rotational invariance, but the effect in this case is negligible.  We
are not interested in the constant $C$ since the ground state energy is
in large measure an artefact of the cut-off.  Our results for both the
$T=5$ and extrapolated potentials are shown in table~\ref{tab:tbigfit}.

\begin{table}[htb]
\caption{Result of fits for $R\ge6a$}
\begin{tabular}{cllr}
\hline
	&\multicolumn{1}{c}{$E$}
	&\multicolumn{1}{c}{$Ka^2$}
	&\multicolumn{1}{c}{$\chi^2/{\rm d.o.f.}$} \\
\hline
	$V(R,5)$ & $0.247(7)$ & $0.00401(8)$ & $15.2/7$ \\
	$V_\infty(R)$ & $0.247(14)$ & $0.00363(17)$ & $8.2/7$ \\
\hline
\end{tabular}
\label{tab:tbigfit}
\end{table}

As already mentioned, we use $V_\infty$ to establish a systematic error.
Hence the value we use for the string tension is from $V(R,5)$:
$Ka^2=0.00401(8)(38)$.

\subsection{Small $R$}
Here our fit includes all separations $R>a$.  Furthermore, we also include
off-axis potentials --- that is, those with separation vectors such as
$(1,1,0)$, $(1,1,1)$, and so forth.  This makes some means of
controlling lattice artefacts imperative.  Our fit in this case has the
form
\begin{equation}
	V(R) = C - {A\over R} + {B\over R^2} + KR - Af\,\delta G(R),
\end{equation}
in which $K$ is fixed from the preceding fit; $A$, $B$ and $f$ are the
free parameters.

There are two new elements.  In this range of separations we are hoping
to be able to see the coupling run; the term in $B$ is a simple-minded
way of correcting the Coulomb term to account for this over a small
range of $R$.

Both $A/R$ and $B/R^2$ terms describe effects which are smooth and
monotonic in $R$.  By contrast, the final term describes lattice
artefacts: $\delta G(R)$ is the one-gluon correction to the Coulomb
term, so that if $f$ were unity the two terms involving $A$ would
combine to give the first-order-corrected Coulomb term which was used in
the previous fit.  By allowing $f$ to vary we account for the following
observation: that the dominant effect of higher-order corrections is to
rescale the first-order term.

\begin{table}[htb]
\caption{Result of fits for $R>a$ (with $K$ fixed)}
\begin{tabular}{cccc}
\hline
	$A$ & $f$ & $B/a$ & $\chi^2/{\rm d.o.f}$ \\
\hline
	$0.238(7)$ & $0.63(3)$ & $0.050(6)$ & $9.7/14$ \\
\hline
\end{tabular}
\label{tab:bigfit}
\end{table}

The $\chi^2$ of the resulting fit (table~\ref{tab:bigfit}) justifies our
use of this ansatz to correct for lack of rotational invariance.
However, it is difficult to estimate the systematic error one should
attach to this.  We have arbitrarily assigned $10\%$ of the correction
which has been applied.

\section{Scaling}

The running coupling is conventionally defined~\cite{beta27} in terms of
$\pot$ as
\begin{equation}
\label{eq:alpha}
	\alpha_{q\bar q}\left(R_1+R_2\over2\right) = {4\over3}R_1R_2
	{V_c(R_1)-V_c(R_2)\over R_1-R_2},
\end{equation}
in which we use the potentials $V_c(R)$ with the lattice-artefact
correction applied.  This is a convenient quantity to use in
investigating scaling as it is dimensionless.

\begin{figure*}[tb]
\begin{center}
\setlength{\unitlength}{0.240900pt}
\ifx\plotpoint\undefined\newsavebox{\plotpoint}\fi
\sbox{\plotpoint}{\rule[-0.175pt]{0.350pt}{0.350pt}}%
\begin{picture}(1500,900)(0,0)
\tenrm
\sbox{\plotpoint}{\rule[-0.175pt]{0.350pt}{0.350pt}}%
\put(264,158){\rule[-0.175pt]{282.335pt}{0.350pt}}
\put(264,158){\rule[-0.175pt]{0.350pt}{151.526pt}}
\put(264,158){\rule[-0.175pt]{4.818pt}{0.350pt}}
\put(242,158){\makebox(0,0)[r]{0}}
\put(1416,158){\rule[-0.175pt]{4.818pt}{0.350pt}}
\put(264,272){\rule[-0.175pt]{4.818pt}{0.350pt}}
\put(242,272){\makebox(0,0)[r]{0.1}}
\put(1416,272){\rule[-0.175pt]{4.818pt}{0.350pt}}
\put(264,387){\rule[-0.175pt]{4.818pt}{0.350pt}}
\put(242,387){\makebox(0,0)[r]{0.2}}
\put(1416,387){\rule[-0.175pt]{4.818pt}{0.350pt}}
\put(264,501){\rule[-0.175pt]{4.818pt}{0.350pt}}
\put(242,501){\makebox(0,0)[r]{0.3}}
\put(1416,501){\rule[-0.175pt]{4.818pt}{0.350pt}}
\put(264,615){\rule[-0.175pt]{4.818pt}{0.350pt}}
\put(242,615){\makebox(0,0)[r]{0.4}}
\put(1416,615){\rule[-0.175pt]{4.818pt}{0.350pt}}
\put(264,730){\rule[-0.175pt]{4.818pt}{0.350pt}}
\put(242,730){\makebox(0,0)[r]{0.5}}
\put(1416,730){\rule[-0.175pt]{4.818pt}{0.350pt}}
\put(264,158){\rule[-0.175pt]{0.350pt}{4.818pt}}
\put(264,113){\makebox(0,0){0}}
\put(264,767){\rule[-0.175pt]{0.350pt}{4.818pt}}
\put(411,158){\rule[-0.175pt]{0.350pt}{4.818pt}}
\put(411,113){\makebox(0,0){0.05}}
\put(411,767){\rule[-0.175pt]{0.350pt}{4.818pt}}
\put(557,158){\rule[-0.175pt]{0.350pt}{4.818pt}}
\put(557,113){\makebox(0,0){0.1}}
\put(557,767){\rule[-0.175pt]{0.350pt}{4.818pt}}
\put(704,158){\rule[-0.175pt]{0.350pt}{4.818pt}}
\put(704,113){\makebox(0,0){0.15}}
\put(704,767){\rule[-0.175pt]{0.350pt}{4.818pt}}
\put(850,158){\rule[-0.175pt]{0.350pt}{4.818pt}}
\put(850,113){\makebox(0,0){0.2}}
\put(850,767){\rule[-0.175pt]{0.350pt}{4.818pt}}
\put(997,158){\rule[-0.175pt]{0.350pt}{4.818pt}}
\put(997,113){\makebox(0,0){0.25}}
\put(997,767){\rule[-0.175pt]{0.350pt}{4.818pt}}
\put(1143,158){\rule[-0.175pt]{0.350pt}{4.818pt}}
\put(1143,113){\makebox(0,0){0.3}}
\put(1143,767){\rule[-0.175pt]{0.350pt}{4.818pt}}
\put(1290,158){\rule[-0.175pt]{0.350pt}{4.818pt}}
\put(1290,113){\makebox(0,0){0.35}}
\put(1290,767){\rule[-0.175pt]{0.350pt}{4.818pt}}
\put(1436,158){\rule[-0.175pt]{0.350pt}{4.818pt}}
\put(1436,113){\makebox(0,0){0.4}}
\put(1436,767){\rule[-0.175pt]{0.350pt}{4.818pt}}
\put(264,158){\rule[-0.175pt]{282.335pt}{0.350pt}}
\put(1436,158){\rule[-0.175pt]{0.350pt}{151.526pt}}
\put(264,787){\rule[-0.175pt]{282.335pt}{0.350pt}}
\put(45,472){\makebox(0,0)[l]{\shortstack{$\alpha_{q\bar q}$}}}
\put(850,68){\makebox(0,0){$R\sqrt K$}}
\put(264,158){\rule[-0.175pt]{0.350pt}{151.526pt}}
\put(488,411){\raisebox{-1.2pt}{\makebox(0,0){$\Diamond$}}}
\put(556,438){\raisebox{-1.2pt}{\makebox(0,0){$\Diamond$}}}
\put(610,458){\raisebox{-1.2pt}{\makebox(0,0){$\Diamond$}}}
\put(657,474){\raisebox{-1.2pt}{\makebox(0,0){$\Diamond$}}}
\put(699,495){\raisebox{-1.2pt}{\makebox(0,0){$\Diamond$}}}
\put(754,502){\raisebox{-1.2pt}{\makebox(0,0){$\Diamond$}}}
\put(805,517){\raisebox{-1.2pt}{\makebox(0,0){$\Diamond$}}}
\put(836,515){\raisebox{-1.2pt}{\makebox(0,0){$\Diamond$}}}
\put(865,533){\raisebox{-1.2pt}{\makebox(0,0){$\Diamond$}}}
\put(893,587){\raisebox{-1.2pt}{\makebox(0,0){$\Diamond$}}}
\put(920,534){\raisebox{-1.2pt}{\makebox(0,0){$\Diamond$}}}
\put(946,570){\raisebox{-1.2pt}{\makebox(0,0){$\Diamond$}}}
\put(994,600){\raisebox{-1.2pt}{\makebox(0,0){$\Diamond$}}}
\put(1029,583){\raisebox{-1.2pt}{\makebox(0,0){$\Diamond$}}}
\put(1051,664){\raisebox{-1.2pt}{\makebox(0,0){$\Diamond$}}}
\put(1073,606){\raisebox{-1.2pt}{\makebox(0,0){$\Diamond$}}}
\put(1134,635){\raisebox{-1.2pt}{\makebox(0,0){$\Diamond$}}}
\put(488,404){\rule[-0.175pt]{0.350pt}{3.613pt}}
\put(478,404){\rule[-0.175pt]{4.818pt}{0.350pt}}
\put(478,419){\rule[-0.175pt]{4.818pt}{0.350pt}}
\put(556,435){\rule[-0.175pt]{0.350pt}{1.445pt}}
\put(546,435){\rule[-0.175pt]{4.818pt}{0.350pt}}
\put(546,441){\rule[-0.175pt]{4.818pt}{0.350pt}}
\put(610,438){\rule[-0.175pt]{0.350pt}{9.636pt}}
\put(600,438){\rule[-0.175pt]{4.818pt}{0.350pt}}
\put(600,478){\rule[-0.175pt]{4.818pt}{0.350pt}}
\put(657,459){\rule[-0.175pt]{0.350pt}{7.227pt}}
\put(647,459){\rule[-0.175pt]{4.818pt}{0.350pt}}
\put(647,489){\rule[-0.175pt]{4.818pt}{0.350pt}}
\put(699,489){\rule[-0.175pt]{0.350pt}{3.132pt}}
\put(689,489){\rule[-0.175pt]{4.818pt}{0.350pt}}
\put(689,502){\rule[-0.175pt]{4.818pt}{0.350pt}}
\put(754,497){\rule[-0.175pt]{0.350pt}{2.409pt}}
\put(744,497){\rule[-0.175pt]{4.818pt}{0.350pt}}
\put(744,507){\rule[-0.175pt]{4.818pt}{0.350pt}}
\put(805,497){\rule[-0.175pt]{0.350pt}{9.395pt}}
\put(795,497){\rule[-0.175pt]{4.818pt}{0.350pt}}
\put(795,536){\rule[-0.175pt]{4.818pt}{0.350pt}}
\put(836,487){\rule[-0.175pt]{0.350pt}{13.490pt}}
\put(826,487){\rule[-0.175pt]{4.818pt}{0.350pt}}
\put(826,543){\rule[-0.175pt]{4.818pt}{0.350pt}}
\put(865,519){\rule[-0.175pt]{0.350pt}{6.745pt}}
\put(855,519){\rule[-0.175pt]{4.818pt}{0.350pt}}
\put(855,547){\rule[-0.175pt]{4.818pt}{0.350pt}}
\put(893,548){\rule[-0.175pt]{0.350pt}{18.549pt}}
\put(883,548){\rule[-0.175pt]{4.818pt}{0.350pt}}
\put(883,625){\rule[-0.175pt]{4.818pt}{0.350pt}}
\put(920,480){\rule[-0.175pt]{0.350pt}{26.017pt}}
\put(910,480){\rule[-0.175pt]{4.818pt}{0.350pt}}
\put(910,588){\rule[-0.175pt]{4.818pt}{0.350pt}}
\put(946,531){\rule[-0.175pt]{0.350pt}{18.549pt}}
\put(936,531){\rule[-0.175pt]{4.818pt}{0.350pt}}
\put(936,608){\rule[-0.175pt]{4.818pt}{0.350pt}}
\put(994,589){\rule[-0.175pt]{0.350pt}{5.541pt}}
\put(984,589){\rule[-0.175pt]{4.818pt}{0.350pt}}
\put(984,612){\rule[-0.175pt]{4.818pt}{0.350pt}}
\put(1029,494){\rule[-0.175pt]{0.350pt}{43.121pt}}
\put(1019,494){\rule[-0.175pt]{4.818pt}{0.350pt}}
\put(1019,673){\rule[-0.175pt]{4.818pt}{0.350pt}}
\put(1051,596){\rule[-0.175pt]{0.350pt}{32.762pt}}
\put(1041,596){\rule[-0.175pt]{4.818pt}{0.350pt}}
\put(1041,732){\rule[-0.175pt]{4.818pt}{0.350pt}}
\put(1073,587){\rule[-0.175pt]{0.350pt}{9.154pt}}
\put(1063,587){\rule[-0.175pt]{4.818pt}{0.350pt}}
\put(1063,625){\rule[-0.175pt]{4.818pt}{0.350pt}}
\put(1134,619){\rule[-0.175pt]{0.350pt}{7.709pt}}
\put(1124,619){\rule[-0.175pt]{4.818pt}{0.350pt}}
\put(1124,651){\rule[-0.175pt]{4.818pt}{0.350pt}}
\put(276,296){\usebox{\plotpoint}}
\put(276,296){\usebox{\plotpoint}}
\put(277,297){\usebox{\plotpoint}}
\put(278,298){\usebox{\plotpoint}}
\put(279,300){\usebox{\plotpoint}}
\put(280,301){\usebox{\plotpoint}}
\put(281,303){\usebox{\plotpoint}}
\put(282,304){\usebox{\plotpoint}}
\put(283,305){\usebox{\plotpoint}}
\put(284,307){\usebox{\plotpoint}}
\put(285,308){\usebox{\plotpoint}}
\put(286,310){\usebox{\plotpoint}}
\put(287,311){\usebox{\plotpoint}}
\put(288,312){\usebox{\plotpoint}}
\put(288,313){\usebox{\plotpoint}}
\put(289,314){\usebox{\plotpoint}}
\put(290,315){\usebox{\plotpoint}}
\put(291,316){\usebox{\plotpoint}}
\put(292,317){\usebox{\plotpoint}}
\put(293,318){\usebox{\plotpoint}}
\put(294,319){\usebox{\plotpoint}}
\put(295,320){\usebox{\plotpoint}}
\put(296,321){\usebox{\plotpoint}}
\put(297,322){\usebox{\plotpoint}}
\put(298,323){\usebox{\plotpoint}}
\put(300,324){\usebox{\plotpoint}}
\put(301,325){\usebox{\plotpoint}}
\put(302,326){\usebox{\plotpoint}}
\put(303,327){\usebox{\plotpoint}}
\put(304,328){\usebox{\plotpoint}}
\put(306,329){\usebox{\plotpoint}}
\put(307,330){\usebox{\plotpoint}}
\put(308,331){\usebox{\plotpoint}}
\put(309,332){\usebox{\plotpoint}}
\put(311,333){\rule[-0.175pt]{0.361pt}{0.350pt}}
\put(312,334){\rule[-0.175pt]{0.361pt}{0.350pt}}
\put(314,335){\rule[-0.175pt]{0.361pt}{0.350pt}}
\put(315,336){\rule[-0.175pt]{0.361pt}{0.350pt}}
\put(317,337){\rule[-0.175pt]{0.361pt}{0.350pt}}
\put(318,338){\rule[-0.175pt]{0.361pt}{0.350pt}}
\put(320,339){\rule[-0.175pt]{0.361pt}{0.350pt}}
\put(321,340){\rule[-0.175pt]{0.361pt}{0.350pt}}
\put(323,341){\rule[-0.175pt]{0.361pt}{0.350pt}}
\put(324,342){\rule[-0.175pt]{0.361pt}{0.350pt}}
\put(326,343){\rule[-0.175pt]{0.361pt}{0.350pt}}
\put(327,344){\rule[-0.175pt]{0.361pt}{0.350pt}}
\put(329,345){\rule[-0.175pt]{0.361pt}{0.350pt}}
\put(330,346){\rule[-0.175pt]{0.361pt}{0.350pt}}
\put(332,347){\rule[-0.175pt]{0.361pt}{0.350pt}}
\put(333,348){\rule[-0.175pt]{0.361pt}{0.350pt}}
\put(335,349){\rule[-0.175pt]{0.482pt}{0.350pt}}
\put(337,350){\rule[-0.175pt]{0.482pt}{0.350pt}}
\put(339,351){\rule[-0.175pt]{0.482pt}{0.350pt}}
\put(341,352){\rule[-0.175pt]{0.482pt}{0.350pt}}
\put(343,353){\rule[-0.175pt]{0.482pt}{0.350pt}}
\put(345,354){\rule[-0.175pt]{0.482pt}{0.350pt}}
\put(347,355){\rule[-0.175pt]{0.482pt}{0.350pt}}
\put(349,356){\rule[-0.175pt]{0.482pt}{0.350pt}}
\put(351,357){\rule[-0.175pt]{0.482pt}{0.350pt}}
\put(353,358){\rule[-0.175pt]{0.482pt}{0.350pt}}
\put(355,359){\rule[-0.175pt]{0.482pt}{0.350pt}}
\put(357,360){\rule[-0.175pt]{0.482pt}{0.350pt}}
\put(359,361){\rule[-0.175pt]{0.482pt}{0.350pt}}
\put(361,362){\rule[-0.175pt]{0.482pt}{0.350pt}}
\put(363,363){\rule[-0.175pt]{0.482pt}{0.350pt}}
\put(365,364){\rule[-0.175pt]{0.482pt}{0.350pt}}
\put(367,365){\rule[-0.175pt]{0.482pt}{0.350pt}}
\put(369,366){\rule[-0.175pt]{0.482pt}{0.350pt}}
\put(371,367){\rule[-0.175pt]{0.530pt}{0.350pt}}
\put(373,368){\rule[-0.175pt]{0.530pt}{0.350pt}}
\put(375,369){\rule[-0.175pt]{0.530pt}{0.350pt}}
\put(377,370){\rule[-0.175pt]{0.530pt}{0.350pt}}
\put(379,371){\rule[-0.175pt]{0.530pt}{0.350pt}}
\put(382,372){\rule[-0.175pt]{0.578pt}{0.350pt}}
\put(384,373){\rule[-0.175pt]{0.578pt}{0.350pt}}
\put(386,374){\rule[-0.175pt]{0.578pt}{0.350pt}}
\put(389,375){\rule[-0.175pt]{0.578pt}{0.350pt}}
\put(391,376){\rule[-0.175pt]{0.578pt}{0.350pt}}
\put(393,377){\rule[-0.175pt]{0.578pt}{0.350pt}}
\put(396,378){\rule[-0.175pt]{0.578pt}{0.350pt}}
\put(398,379){\rule[-0.175pt]{0.578pt}{0.350pt}}
\put(401,380){\rule[-0.175pt]{0.578pt}{0.350pt}}
\put(403,381){\rule[-0.175pt]{0.578pt}{0.350pt}}
\put(405,382){\rule[-0.175pt]{0.578pt}{0.350pt}}
\put(408,383){\rule[-0.175pt]{0.578pt}{0.350pt}}
\put(410,384){\rule[-0.175pt]{0.578pt}{0.350pt}}
\put(413,385){\rule[-0.175pt]{0.578pt}{0.350pt}}
\put(415,386){\rule[-0.175pt]{0.578pt}{0.350pt}}
\put(417,387){\rule[-0.175pt]{0.723pt}{0.350pt}}
\put(421,388){\rule[-0.175pt]{0.723pt}{0.350pt}}
\put(424,389){\rule[-0.175pt]{0.723pt}{0.350pt}}
\put(427,390){\rule[-0.175pt]{0.723pt}{0.350pt}}
\put(430,391){\rule[-0.175pt]{0.578pt}{0.350pt}}
\put(432,392){\rule[-0.175pt]{0.578pt}{0.350pt}}
\put(434,393){\rule[-0.175pt]{0.578pt}{0.350pt}}
\put(437,394){\rule[-0.175pt]{0.578pt}{0.350pt}}
\put(439,395){\rule[-0.175pt]{0.578pt}{0.350pt}}
\put(441,396){\rule[-0.175pt]{0.662pt}{0.350pt}}
\put(444,397){\rule[-0.175pt]{0.662pt}{0.350pt}}
\put(447,398){\rule[-0.175pt]{0.662pt}{0.350pt}}
\put(450,399){\rule[-0.175pt]{0.662pt}{0.350pt}}
\put(453,400){\rule[-0.175pt]{0.723pt}{0.350pt}}
\put(456,401){\rule[-0.175pt]{0.723pt}{0.350pt}}
\put(459,402){\rule[-0.175pt]{0.723pt}{0.350pt}}
\put(462,403){\rule[-0.175pt]{0.723pt}{0.350pt}}
\put(465,404){\rule[-0.175pt]{0.723pt}{0.350pt}}
\put(468,405){\rule[-0.175pt]{0.723pt}{0.350pt}}
\put(471,406){\rule[-0.175pt]{0.723pt}{0.350pt}}
\put(474,407){\rule[-0.175pt]{0.723pt}{0.350pt}}
\put(477,408){\rule[-0.175pt]{0.723pt}{0.350pt}}
\put(480,409){\rule[-0.175pt]{0.723pt}{0.350pt}}
\put(483,410){\rule[-0.175pt]{0.723pt}{0.350pt}}
\put(486,411){\rule[-0.175pt]{0.723pt}{0.350pt}}
\put(489,412){\rule[-0.175pt]{0.723pt}{0.350pt}}
\put(492,413){\rule[-0.175pt]{0.723pt}{0.350pt}}
\put(495,414){\rule[-0.175pt]{0.723pt}{0.350pt}}
\put(498,415){\rule[-0.175pt]{0.723pt}{0.350pt}}
\put(501,416){\rule[-0.175pt]{0.723pt}{0.350pt}}
\put(504,417){\rule[-0.175pt]{0.723pt}{0.350pt}}
\put(507,418){\rule[-0.175pt]{0.723pt}{0.350pt}}
\put(510,419){\rule[-0.175pt]{0.723pt}{0.350pt}}
\put(513,420){\rule[-0.175pt]{0.662pt}{0.350pt}}
\put(515,421){\rule[-0.175pt]{0.662pt}{0.350pt}}
\put(518,422){\rule[-0.175pt]{0.662pt}{0.350pt}}
\put(521,423){\rule[-0.175pt]{0.662pt}{0.350pt}}
\put(524,424){\rule[-0.175pt]{0.964pt}{0.350pt}}
\put(528,425){\rule[-0.175pt]{0.964pt}{0.350pt}}
\put(532,426){\rule[-0.175pt]{0.964pt}{0.350pt}}
\put(536,427){\rule[-0.175pt]{0.723pt}{0.350pt}}
\put(539,428){\rule[-0.175pt]{0.723pt}{0.350pt}}
\put(542,429){\rule[-0.175pt]{0.723pt}{0.350pt}}
\put(545,430){\rule[-0.175pt]{0.723pt}{0.350pt}}
\put(548,431){\rule[-0.175pt]{0.723pt}{0.350pt}}
\put(551,432){\rule[-0.175pt]{0.723pt}{0.350pt}}
\put(554,433){\rule[-0.175pt]{0.723pt}{0.350pt}}
\put(557,434){\rule[-0.175pt]{0.723pt}{0.350pt}}
\put(560,435){\rule[-0.175pt]{0.964pt}{0.350pt}}
\put(564,436){\rule[-0.175pt]{0.964pt}{0.350pt}}
\put(568,437){\rule[-0.175pt]{0.964pt}{0.350pt}}
\put(572,438){\rule[-0.175pt]{0.723pt}{0.350pt}}
\put(575,439){\rule[-0.175pt]{0.723pt}{0.350pt}}
\put(578,440){\rule[-0.175pt]{0.723pt}{0.350pt}}
\put(581,441){\rule[-0.175pt]{0.723pt}{0.350pt}}
\put(584,442){\rule[-0.175pt]{0.883pt}{0.350pt}}
\put(587,443){\rule[-0.175pt]{0.883pt}{0.350pt}}
\put(591,444){\rule[-0.175pt]{0.883pt}{0.350pt}}
\put(595,445){\rule[-0.175pt]{0.723pt}{0.350pt}}
\put(598,446){\rule[-0.175pt]{0.723pt}{0.350pt}}
\put(601,447){\rule[-0.175pt]{0.723pt}{0.350pt}}
\put(604,448){\rule[-0.175pt]{0.723pt}{0.350pt}}
\put(607,449){\rule[-0.175pt]{0.964pt}{0.350pt}}
\put(611,450){\rule[-0.175pt]{0.964pt}{0.350pt}}
\put(615,451){\rule[-0.175pt]{0.964pt}{0.350pt}}
\put(619,452){\rule[-0.175pt]{0.964pt}{0.350pt}}
\put(623,453){\rule[-0.175pt]{0.964pt}{0.350pt}}
\put(627,454){\rule[-0.175pt]{0.964pt}{0.350pt}}
\put(631,455){\rule[-0.175pt]{0.723pt}{0.350pt}}
\put(634,456){\rule[-0.175pt]{0.723pt}{0.350pt}}
\put(637,457){\rule[-0.175pt]{0.723pt}{0.350pt}}
\put(640,458){\rule[-0.175pt]{0.723pt}{0.350pt}}
\put(643,459){\rule[-0.175pt]{0.964pt}{0.350pt}}
\put(647,460){\rule[-0.175pt]{0.964pt}{0.350pt}}
\put(651,461){\rule[-0.175pt]{0.964pt}{0.350pt}}
\put(655,462){\rule[-0.175pt]{0.723pt}{0.350pt}}
\put(658,463){\rule[-0.175pt]{0.723pt}{0.350pt}}
\put(661,464){\rule[-0.175pt]{0.723pt}{0.350pt}}
\put(664,465){\rule[-0.175pt]{0.723pt}{0.350pt}}
\put(667,466){\rule[-0.175pt]{0.883pt}{0.350pt}}
\put(670,467){\rule[-0.175pt]{0.883pt}{0.350pt}}
\put(674,468){\rule[-0.175pt]{0.883pt}{0.350pt}}
\put(678,469){\rule[-0.175pt]{0.964pt}{0.350pt}}
\put(682,470){\rule[-0.175pt]{0.964pt}{0.350pt}}
\put(686,471){\rule[-0.175pt]{0.964pt}{0.350pt}}
\put(690,472){\rule[-0.175pt]{0.964pt}{0.350pt}}
\put(694,473){\rule[-0.175pt]{0.964pt}{0.350pt}}
\put(698,474){\rule[-0.175pt]{0.964pt}{0.350pt}}
\put(702,475){\rule[-0.175pt]{0.723pt}{0.350pt}}
\put(705,476){\rule[-0.175pt]{0.723pt}{0.350pt}}
\put(708,477){\rule[-0.175pt]{0.723pt}{0.350pt}}
\put(711,478){\rule[-0.175pt]{0.723pt}{0.350pt}}
\put(714,479){\rule[-0.175pt]{0.964pt}{0.350pt}}
\put(718,480){\rule[-0.175pt]{0.964pt}{0.350pt}}
\put(722,481){\rule[-0.175pt]{0.964pt}{0.350pt}}
\put(726,482){\rule[-0.175pt]{0.964pt}{0.350pt}}
\put(730,483){\rule[-0.175pt]{0.964pt}{0.350pt}}
\put(734,484){\rule[-0.175pt]{0.964pt}{0.350pt}}
\put(738,485){\rule[-0.175pt]{0.883pt}{0.350pt}}
\put(741,486){\rule[-0.175pt]{0.883pt}{0.350pt}}
\put(745,487){\rule[-0.175pt]{0.883pt}{0.350pt}}
\put(749,488){\rule[-0.175pt]{0.723pt}{0.350pt}}
\put(752,489){\rule[-0.175pt]{0.723pt}{0.350pt}}
\put(755,490){\rule[-0.175pt]{0.723pt}{0.350pt}}
\put(758,491){\rule[-0.175pt]{0.723pt}{0.350pt}}
\put(761,492){\rule[-0.175pt]{0.964pt}{0.350pt}}
\put(765,493){\rule[-0.175pt]{0.964pt}{0.350pt}}
\put(769,494){\rule[-0.175pt]{0.964pt}{0.350pt}}
\put(773,495){\rule[-0.175pt]{0.964pt}{0.350pt}}
\put(777,496){\rule[-0.175pt]{0.964pt}{0.350pt}}
\put(781,497){\rule[-0.175pt]{0.964pt}{0.350pt}}
\put(785,498){\rule[-0.175pt]{0.964pt}{0.350pt}}
\put(789,499){\rule[-0.175pt]{0.964pt}{0.350pt}}
\put(793,500){\rule[-0.175pt]{0.964pt}{0.350pt}}
\put(797,501){\rule[-0.175pt]{0.964pt}{0.350pt}}
\put(801,502){\rule[-0.175pt]{0.964pt}{0.350pt}}
\put(805,503){\rule[-0.175pt]{0.964pt}{0.350pt}}
\put(809,504){\rule[-0.175pt]{0.883pt}{0.350pt}}
\put(812,505){\rule[-0.175pt]{0.883pt}{0.350pt}}
\put(816,506){\rule[-0.175pt]{0.883pt}{0.350pt}}
\put(820,507){\rule[-0.175pt]{0.723pt}{0.350pt}}
\put(823,508){\rule[-0.175pt]{0.723pt}{0.350pt}}
\put(826,509){\rule[-0.175pt]{0.723pt}{0.350pt}}
\put(829,510){\rule[-0.175pt]{0.723pt}{0.350pt}}
\put(832,511){\rule[-0.175pt]{0.964pt}{0.350pt}}
\put(836,512){\rule[-0.175pt]{0.964pt}{0.350pt}}
\put(840,513){\rule[-0.175pt]{0.964pt}{0.350pt}}
\put(844,514){\rule[-0.175pt]{0.964pt}{0.350pt}}
\put(848,515){\rule[-0.175pt]{0.964pt}{0.350pt}}
\put(852,516){\rule[-0.175pt]{0.964pt}{0.350pt}}
\put(856,517){\rule[-0.175pt]{0.964pt}{0.350pt}}
\put(860,518){\rule[-0.175pt]{0.964pt}{0.350pt}}
\put(864,519){\rule[-0.175pt]{0.964pt}{0.350pt}}
\put(868,520){\rule[-0.175pt]{0.964pt}{0.350pt}}
\put(872,521){\rule[-0.175pt]{0.964pt}{0.350pt}}
\put(876,522){\rule[-0.175pt]{0.964pt}{0.350pt}}
\put(880,523){\rule[-0.175pt]{0.883pt}{0.350pt}}
\put(883,524){\rule[-0.175pt]{0.883pt}{0.350pt}}
\put(887,525){\rule[-0.175pt]{0.883pt}{0.350pt}}
\put(891,526){\rule[-0.175pt]{0.964pt}{0.350pt}}
\put(895,527){\rule[-0.175pt]{0.964pt}{0.350pt}}
\put(899,528){\rule[-0.175pt]{0.964pt}{0.350pt}}
\put(903,529){\rule[-0.175pt]{0.723pt}{0.350pt}}
\put(906,530){\rule[-0.175pt]{0.723pt}{0.350pt}}
\put(909,531){\rule[-0.175pt]{0.723pt}{0.350pt}}
\put(912,532){\rule[-0.175pt]{0.723pt}{0.350pt}}
\put(915,533){\rule[-0.175pt]{0.964pt}{0.350pt}}
\put(919,534){\rule[-0.175pt]{0.964pt}{0.350pt}}
\put(923,535){\rule[-0.175pt]{0.964pt}{0.350pt}}
\put(927,536){\rule[-0.175pt]{0.964pt}{0.350pt}}
\put(931,537){\rule[-0.175pt]{0.964pt}{0.350pt}}
\put(935,538){\rule[-0.175pt]{0.964pt}{0.350pt}}
\put(939,539){\rule[-0.175pt]{0.964pt}{0.350pt}}
\put(943,540){\rule[-0.175pt]{0.964pt}{0.350pt}}
\put(947,541){\rule[-0.175pt]{0.964pt}{0.350pt}}
\put(951,542){\rule[-0.175pt]{0.883pt}{0.350pt}}
\put(954,543){\rule[-0.175pt]{0.883pt}{0.350pt}}
\put(958,544){\rule[-0.175pt]{0.883pt}{0.350pt}}
\put(962,545){\rule[-0.175pt]{0.964pt}{0.350pt}}
\put(966,546){\rule[-0.175pt]{0.964pt}{0.350pt}}
\put(970,547){\rule[-0.175pt]{0.964pt}{0.350pt}}
\put(974,548){\rule[-0.175pt]{0.723pt}{0.350pt}}
\put(977,549){\rule[-0.175pt]{0.723pt}{0.350pt}}
\put(980,550){\rule[-0.175pt]{0.723pt}{0.350pt}}
\put(983,551){\rule[-0.175pt]{0.723pt}{0.350pt}}
\put(986,552){\rule[-0.175pt]{0.964pt}{0.350pt}}
\put(990,553){\rule[-0.175pt]{0.964pt}{0.350pt}}
\put(994,554){\rule[-0.175pt]{0.964pt}{0.350pt}}
\put(998,555){\rule[-0.175pt]{0.964pt}{0.350pt}}
\put(1002,556){\rule[-0.175pt]{0.964pt}{0.350pt}}
\put(1006,557){\rule[-0.175pt]{0.964pt}{0.350pt}}
\put(1010,558){\rule[-0.175pt]{0.964pt}{0.350pt}}
\put(1014,559){\rule[-0.175pt]{0.964pt}{0.350pt}}
\put(1018,560){\rule[-0.175pt]{0.964pt}{0.350pt}}
\put(1022,561){\rule[-0.175pt]{0.883pt}{0.350pt}}
\put(1025,562){\rule[-0.175pt]{0.883pt}{0.350pt}}
\put(1029,563){\rule[-0.175pt]{0.883pt}{0.350pt}}
\put(1032,564){\rule[-0.175pt]{0.964pt}{0.350pt}}
\put(1037,565){\rule[-0.175pt]{0.964pt}{0.350pt}}
\put(1041,566){\rule[-0.175pt]{0.964pt}{0.350pt}}
\put(1045,567){\rule[-0.175pt]{0.723pt}{0.350pt}}
\put(1048,568){\rule[-0.175pt]{0.723pt}{0.350pt}}
\put(1051,569){\rule[-0.175pt]{0.723pt}{0.350pt}}
\put(1054,570){\rule[-0.175pt]{0.723pt}{0.350pt}}
\put(1057,571){\rule[-0.175pt]{0.964pt}{0.350pt}}
\put(1061,572){\rule[-0.175pt]{0.964pt}{0.350pt}}
\put(1065,573){\rule[-0.175pt]{0.964pt}{0.350pt}}
\put(1069,574){\rule[-0.175pt]{0.964pt}{0.350pt}}
\put(1073,575){\rule[-0.175pt]{0.964pt}{0.350pt}}
\put(1077,576){\rule[-0.175pt]{0.964pt}{0.350pt}}
\put(1081,577){\rule[-0.175pt]{0.964pt}{0.350pt}}
\put(1085,578){\rule[-0.175pt]{0.964pt}{0.350pt}}
\put(1089,579){\rule[-0.175pt]{0.964pt}{0.350pt}}
\put(1093,580){\rule[-0.175pt]{0.964pt}{0.350pt}}
\put(1097,581){\rule[-0.175pt]{0.964pt}{0.350pt}}
\put(1101,582){\rule[-0.175pt]{0.964pt}{0.350pt}}
\put(1105,583){\rule[-0.175pt]{0.662pt}{0.350pt}}
\put(1107,584){\rule[-0.175pt]{0.662pt}{0.350pt}}
\put(1110,585){\rule[-0.175pt]{0.662pt}{0.350pt}}
\put(1113,586){\rule[-0.175pt]{0.662pt}{0.350pt}}
\put(1116,587){\rule[-0.175pt]{0.964pt}{0.350pt}}
\put(1120,588){\rule[-0.175pt]{0.964pt}{0.350pt}}
\put(1124,589){\rule[-0.175pt]{0.964pt}{0.350pt}}
\put(1128,590){\rule[-0.175pt]{0.964pt}{0.350pt}}
\put(1132,591){\rule[-0.175pt]{0.964pt}{0.350pt}}
\put(1136,592){\rule[-0.175pt]{0.964pt}{0.350pt}}
\put(1140,593){\rule[-0.175pt]{0.964pt}{0.350pt}}
\put(1144,594){\rule[-0.175pt]{0.964pt}{0.350pt}}
\put(1148,595){\rule[-0.175pt]{0.964pt}{0.350pt}}
\put(1152,596){\rule[-0.175pt]{0.723pt}{0.350pt}}
\put(1155,597){\rule[-0.175pt]{0.723pt}{0.350pt}}
\put(1158,598){\rule[-0.175pt]{0.723pt}{0.350pt}}
\put(1161,599){\rule[-0.175pt]{0.723pt}{0.350pt}}
\put(1164,600){\rule[-0.175pt]{0.964pt}{0.350pt}}
\put(1168,601){\rule[-0.175pt]{0.964pt}{0.350pt}}
\put(1172,602){\rule[-0.175pt]{0.964pt}{0.350pt}}
\put(1176,603){\rule[-0.175pt]{0.883pt}{0.350pt}}
\put(1179,604){\rule[-0.175pt]{0.883pt}{0.350pt}}
\put(1183,605){\rule[-0.175pt]{0.883pt}{0.350pt}}
\put(1186,606){\rule[-0.175pt]{0.964pt}{0.350pt}}
\put(1191,607){\rule[-0.175pt]{0.964pt}{0.350pt}}
\put(1195,608){\rule[-0.175pt]{0.964pt}{0.350pt}}
\put(1199,609){\rule[-0.175pt]{0.723pt}{0.350pt}}
\put(1202,610){\rule[-0.175pt]{0.723pt}{0.350pt}}
\put(1205,611){\rule[-0.175pt]{0.723pt}{0.350pt}}
\put(1208,612){\rule[-0.175pt]{0.723pt}{0.350pt}}
\put(1211,613){\rule[-0.175pt]{0.964pt}{0.350pt}}
\put(1215,614){\rule[-0.175pt]{0.964pt}{0.350pt}}
\put(1219,615){\rule[-0.175pt]{0.964pt}{0.350pt}}
\put(1223,616){\rule[-0.175pt]{0.964pt}{0.350pt}}
\put(1227,617){\rule[-0.175pt]{0.964pt}{0.350pt}}
\put(1231,618){\rule[-0.175pt]{0.964pt}{0.350pt}}
\put(1235,619){\rule[-0.175pt]{0.723pt}{0.350pt}}
\put(1238,620){\rule[-0.175pt]{0.723pt}{0.350pt}}
\put(1241,621){\rule[-0.175pt]{0.723pt}{0.350pt}}
\put(1244,622){\rule[-0.175pt]{0.723pt}{0.350pt}}
\put(1247,623){\rule[-0.175pt]{0.883pt}{0.350pt}}
\put(1250,624){\rule[-0.175pt]{0.883pt}{0.350pt}}
\put(1254,625){\rule[-0.175pt]{0.883pt}{0.350pt}}
\put(1257,626){\rule[-0.175pt]{0.964pt}{0.350pt}}
\put(1262,627){\rule[-0.175pt]{0.964pt}{0.350pt}}
\put(1266,628){\rule[-0.175pt]{0.964pt}{0.350pt}}
\put(1270,629){\rule[-0.175pt]{0.723pt}{0.350pt}}
\put(1273,630){\rule[-0.175pt]{0.723pt}{0.350pt}}
\put(1276,631){\rule[-0.175pt]{0.723pt}{0.350pt}}
\put(1279,632){\rule[-0.175pt]{0.723pt}{0.350pt}}
\put(1282,633){\rule[-0.175pt]{0.964pt}{0.350pt}}
\put(1286,634){\rule[-0.175pt]{0.964pt}{0.350pt}}
\put(1290,635){\rule[-0.175pt]{0.964pt}{0.350pt}}
\put(1294,636){\rule[-0.175pt]{0.723pt}{0.350pt}}
\put(1297,637){\rule[-0.175pt]{0.723pt}{0.350pt}}
\put(1300,638){\rule[-0.175pt]{0.723pt}{0.350pt}}
\put(1303,639){\rule[-0.175pt]{0.723pt}{0.350pt}}
\put(1306,640){\rule[-0.175pt]{0.964pt}{0.350pt}}
\put(1310,641){\rule[-0.175pt]{0.964pt}{0.350pt}}
\put(1314,642){\rule[-0.175pt]{0.964pt}{0.350pt}}
\put(1318,643){\rule[-0.175pt]{0.662pt}{0.350pt}}
\put(1320,644){\rule[-0.175pt]{0.662pt}{0.350pt}}
\put(1323,645){\rule[-0.175pt]{0.662pt}{0.350pt}}
\put(1326,646){\rule[-0.175pt]{0.662pt}{0.350pt}}
\put(1329,647){\rule[-0.175pt]{0.964pt}{0.350pt}}
\put(1333,648){\rule[-0.175pt]{0.964pt}{0.350pt}}
\put(1337,649){\rule[-0.175pt]{0.964pt}{0.350pt}}
\put(1341,650){\rule[-0.175pt]{0.964pt}{0.350pt}}
\put(1345,651){\rule[-0.175pt]{0.964pt}{0.350pt}}
\put(1349,652){\rule[-0.175pt]{0.964pt}{0.350pt}}
\put(1353,653){\rule[-0.175pt]{0.723pt}{0.350pt}}
\put(1356,654){\rule[-0.175pt]{0.723pt}{0.350pt}}
\put(1359,655){\rule[-0.175pt]{0.723pt}{0.350pt}}
\put(1362,656){\rule[-0.175pt]{0.723pt}{0.350pt}}
\put(1365,657){\rule[-0.175pt]{0.964pt}{0.350pt}}
\put(1369,658){\rule[-0.175pt]{0.964pt}{0.350pt}}
\put(1373,659){\rule[-0.175pt]{0.964pt}{0.350pt}}
\put(1377,660){\rule[-0.175pt]{0.723pt}{0.350pt}}
\put(1380,661){\rule[-0.175pt]{0.723pt}{0.350pt}}
\put(1383,662){\rule[-0.175pt]{0.723pt}{0.350pt}}
\put(1386,663){\rule[-0.175pt]{0.723pt}{0.350pt}}
\put(1389,664){\rule[-0.175pt]{0.662pt}{0.350pt}}
\put(1391,665){\rule[-0.175pt]{0.662pt}{0.350pt}}
\put(1394,666){\rule[-0.175pt]{0.662pt}{0.350pt}}
\put(1397,667){\rule[-0.175pt]{0.662pt}{0.350pt}}
\put(1400,668){\rule[-0.175pt]{0.964pt}{0.350pt}}
\put(1404,669){\rule[-0.175pt]{0.964pt}{0.350pt}}
\put(1408,670){\rule[-0.175pt]{0.964pt}{0.350pt}}
\put(1412,671){\rule[-0.175pt]{0.723pt}{0.350pt}}
\put(1415,672){\rule[-0.175pt]{0.723pt}{0.350pt}}
\put(1418,673){\rule[-0.175pt]{0.723pt}{0.350pt}}
\put(1421,674){\rule[-0.175pt]{0.723pt}{0.350pt}}
\put(1424,675){\rule[-0.175pt]{0.964pt}{0.350pt}}
\put(1428,676){\rule[-0.175pt]{0.964pt}{0.350pt}}
\put(1432,677){\rule[-0.175pt]{0.964pt}{0.350pt}}
\put(276,299){\usebox{\plotpoint}}
\put(276,299){\rule[-0.175pt]{0.350pt}{0.361pt}}
\put(277,300){\rule[-0.175pt]{0.350pt}{0.361pt}}
\put(278,302){\rule[-0.175pt]{0.350pt}{0.361pt}}
\put(279,303){\rule[-0.175pt]{0.350pt}{0.361pt}}
\put(280,305){\rule[-0.175pt]{0.350pt}{0.361pt}}
\put(281,306){\rule[-0.175pt]{0.350pt}{0.361pt}}
\put(282,308){\rule[-0.175pt]{0.350pt}{0.361pt}}
\put(283,309){\rule[-0.175pt]{0.350pt}{0.361pt}}
\put(284,311){\rule[-0.175pt]{0.350pt}{0.361pt}}
\put(285,312){\rule[-0.175pt]{0.350pt}{0.361pt}}
\put(286,314){\rule[-0.175pt]{0.350pt}{0.361pt}}
\put(287,315){\rule[-0.175pt]{0.350pt}{0.361pt}}
\put(288,317){\usebox{\plotpoint}}
\put(289,318){\usebox{\plotpoint}}
\put(290,319){\usebox{\plotpoint}}
\put(291,320){\usebox{\plotpoint}}
\put(292,321){\usebox{\plotpoint}}
\put(293,322){\usebox{\plotpoint}}
\put(294,323){\usebox{\plotpoint}}
\put(295,324){\usebox{\plotpoint}}
\put(296,325){\usebox{\plotpoint}}
\put(297,326){\usebox{\plotpoint}}
\put(298,327){\usebox{\plotpoint}}
\put(299,328){\usebox{\plotpoint}}
\put(300,329){\usebox{\plotpoint}}
\put(301,330){\usebox{\plotpoint}}
\put(302,331){\usebox{\plotpoint}}
\put(303,332){\usebox{\plotpoint}}
\put(304,333){\usebox{\plotpoint}}
\put(305,334){\usebox{\plotpoint}}
\put(306,335){\usebox{\plotpoint}}
\put(307,336){\usebox{\plotpoint}}
\put(308,337){\usebox{\plotpoint}}
\put(309,338){\usebox{\plotpoint}}
\put(311,339){\rule[-0.175pt]{0.361pt}{0.350pt}}
\put(312,340){\rule[-0.175pt]{0.361pt}{0.350pt}}
\put(314,341){\rule[-0.175pt]{0.361pt}{0.350pt}}
\put(315,342){\rule[-0.175pt]{0.361pt}{0.350pt}}
\put(317,343){\rule[-0.175pt]{0.361pt}{0.350pt}}
\put(318,344){\rule[-0.175pt]{0.361pt}{0.350pt}}
\put(320,345){\rule[-0.175pt]{0.361pt}{0.350pt}}
\put(321,346){\rule[-0.175pt]{0.361pt}{0.350pt}}
\put(323,347){\rule[-0.175pt]{0.361pt}{0.350pt}}
\put(324,348){\rule[-0.175pt]{0.361pt}{0.350pt}}
\put(326,349){\rule[-0.175pt]{0.361pt}{0.350pt}}
\put(327,350){\rule[-0.175pt]{0.361pt}{0.350pt}}
\put(329,351){\rule[-0.175pt]{0.361pt}{0.350pt}}
\put(330,352){\rule[-0.175pt]{0.361pt}{0.350pt}}
\put(332,353){\rule[-0.175pt]{0.361pt}{0.350pt}}
\put(333,354){\rule[-0.175pt]{0.361pt}{0.350pt}}
\put(335,355){\rule[-0.175pt]{0.413pt}{0.350pt}}
\put(336,356){\rule[-0.175pt]{0.413pt}{0.350pt}}
\put(338,357){\rule[-0.175pt]{0.413pt}{0.350pt}}
\put(340,358){\rule[-0.175pt]{0.413pt}{0.350pt}}
\put(341,359){\rule[-0.175pt]{0.413pt}{0.350pt}}
\put(343,360){\rule[-0.175pt]{0.413pt}{0.350pt}}
\put(345,361){\rule[-0.175pt]{0.413pt}{0.350pt}}
\put(347,362){\rule[-0.175pt]{0.482pt}{0.350pt}}
\put(349,363){\rule[-0.175pt]{0.482pt}{0.350pt}}
\put(351,364){\rule[-0.175pt]{0.482pt}{0.350pt}}
\put(353,365){\rule[-0.175pt]{0.482pt}{0.350pt}}
\put(355,366){\rule[-0.175pt]{0.482pt}{0.350pt}}
\put(357,367){\rule[-0.175pt]{0.482pt}{0.350pt}}
\put(359,368){\rule[-0.175pt]{0.482pt}{0.350pt}}
\put(361,369){\rule[-0.175pt]{0.482pt}{0.350pt}}
\put(363,370){\rule[-0.175pt]{0.482pt}{0.350pt}}
\put(365,371){\rule[-0.175pt]{0.482pt}{0.350pt}}
\put(367,372){\rule[-0.175pt]{0.482pt}{0.350pt}}
\put(369,373){\rule[-0.175pt]{0.482pt}{0.350pt}}
\put(371,374){\rule[-0.175pt]{0.442pt}{0.350pt}}
\put(372,375){\rule[-0.175pt]{0.442pt}{0.350pt}}
\put(374,376){\rule[-0.175pt]{0.442pt}{0.350pt}}
\put(376,377){\rule[-0.175pt]{0.442pt}{0.350pt}}
\put(378,378){\rule[-0.175pt]{0.442pt}{0.350pt}}
\put(380,379){\rule[-0.175pt]{0.442pt}{0.350pt}}
\put(382,380){\rule[-0.175pt]{0.578pt}{0.350pt}}
\put(384,381){\rule[-0.175pt]{0.578pt}{0.350pt}}
\put(386,382){\rule[-0.175pt]{0.578pt}{0.350pt}}
\put(389,383){\rule[-0.175pt]{0.578pt}{0.350pt}}
\put(391,384){\rule[-0.175pt]{0.578pt}{0.350pt}}
\put(393,385){\rule[-0.175pt]{0.482pt}{0.350pt}}
\put(396,386){\rule[-0.175pt]{0.482pt}{0.350pt}}
\put(398,387){\rule[-0.175pt]{0.482pt}{0.350pt}}
\put(400,388){\rule[-0.175pt]{0.482pt}{0.350pt}}
\put(402,389){\rule[-0.175pt]{0.482pt}{0.350pt}}
\put(404,390){\rule[-0.175pt]{0.482pt}{0.350pt}}
\put(406,391){\rule[-0.175pt]{0.578pt}{0.350pt}}
\put(408,392){\rule[-0.175pt]{0.578pt}{0.350pt}}
\put(410,393){\rule[-0.175pt]{0.578pt}{0.350pt}}
\put(413,394){\rule[-0.175pt]{0.578pt}{0.350pt}}
\put(415,395){\rule[-0.175pt]{0.578pt}{0.350pt}}
\put(417,396){\rule[-0.175pt]{0.578pt}{0.350pt}}
\put(420,397){\rule[-0.175pt]{0.578pt}{0.350pt}}
\put(422,398){\rule[-0.175pt]{0.578pt}{0.350pt}}
\put(425,399){\rule[-0.175pt]{0.578pt}{0.350pt}}
\put(427,400){\rule[-0.175pt]{0.578pt}{0.350pt}}
\put(429,401){\rule[-0.175pt]{0.578pt}{0.350pt}}
\put(432,402){\rule[-0.175pt]{0.578pt}{0.350pt}}
\put(434,403){\rule[-0.175pt]{0.578pt}{0.350pt}}
\put(437,404){\rule[-0.175pt]{0.578pt}{0.350pt}}
\put(439,405){\rule[-0.175pt]{0.578pt}{0.350pt}}
\put(441,406){\rule[-0.175pt]{0.662pt}{0.350pt}}
\put(444,407){\rule[-0.175pt]{0.662pt}{0.350pt}}
\put(447,408){\rule[-0.175pt]{0.662pt}{0.350pt}}
\put(450,409){\rule[-0.175pt]{0.662pt}{0.350pt}}
\put(453,410){\rule[-0.175pt]{0.578pt}{0.350pt}}
\put(455,411){\rule[-0.175pt]{0.578pt}{0.350pt}}
\put(457,412){\rule[-0.175pt]{0.578pt}{0.350pt}}
\put(460,413){\rule[-0.175pt]{0.578pt}{0.350pt}}
\put(462,414){\rule[-0.175pt]{0.578pt}{0.350pt}}
\put(464,415){\rule[-0.175pt]{0.723pt}{0.350pt}}
\put(468,416){\rule[-0.175pt]{0.723pt}{0.350pt}}
\put(471,417){\rule[-0.175pt]{0.723pt}{0.350pt}}
\put(474,418){\rule[-0.175pt]{0.723pt}{0.350pt}}
\put(477,419){\rule[-0.175pt]{0.578pt}{0.350pt}}
\put(479,420){\rule[-0.175pt]{0.578pt}{0.350pt}}
\put(481,421){\rule[-0.175pt]{0.578pt}{0.350pt}}
\put(484,422){\rule[-0.175pt]{0.578pt}{0.350pt}}
\put(486,423){\rule[-0.175pt]{0.578pt}{0.350pt}}
\put(488,424){\rule[-0.175pt]{0.723pt}{0.350pt}}
\put(492,425){\rule[-0.175pt]{0.723pt}{0.350pt}}
\put(495,426){\rule[-0.175pt]{0.723pt}{0.350pt}}
\put(498,427){\rule[-0.175pt]{0.723pt}{0.350pt}}
\put(501,428){\rule[-0.175pt]{0.723pt}{0.350pt}}
\put(504,429){\rule[-0.175pt]{0.723pt}{0.350pt}}
\put(507,430){\rule[-0.175pt]{0.723pt}{0.350pt}}
\put(510,431){\rule[-0.175pt]{0.723pt}{0.350pt}}
\put(513,432){\rule[-0.175pt]{0.662pt}{0.350pt}}
\put(515,433){\rule[-0.175pt]{0.662pt}{0.350pt}}
\put(518,434){\rule[-0.175pt]{0.662pt}{0.350pt}}
\put(521,435){\rule[-0.175pt]{0.662pt}{0.350pt}}
\put(524,436){\rule[-0.175pt]{0.723pt}{0.350pt}}
\put(527,437){\rule[-0.175pt]{0.723pt}{0.350pt}}
\put(530,438){\rule[-0.175pt]{0.723pt}{0.350pt}}
\put(533,439){\rule[-0.175pt]{0.723pt}{0.350pt}}
\put(536,440){\rule[-0.175pt]{0.723pt}{0.350pt}}
\put(539,441){\rule[-0.175pt]{0.723pt}{0.350pt}}
\put(542,442){\rule[-0.175pt]{0.723pt}{0.350pt}}
\put(545,443){\rule[-0.175pt]{0.723pt}{0.350pt}}
\put(548,444){\rule[-0.175pt]{0.723pt}{0.350pt}}
\put(551,445){\rule[-0.175pt]{0.723pt}{0.350pt}}
\put(554,446){\rule[-0.175pt]{0.723pt}{0.350pt}}
\put(557,447){\rule[-0.175pt]{0.723pt}{0.350pt}}
\put(560,448){\rule[-0.175pt]{0.723pt}{0.350pt}}
\put(563,449){\rule[-0.175pt]{0.723pt}{0.350pt}}
\put(566,450){\rule[-0.175pt]{0.723pt}{0.350pt}}
\put(569,451){\rule[-0.175pt]{0.723pt}{0.350pt}}
\put(572,452){\rule[-0.175pt]{0.723pt}{0.350pt}}
\put(575,453){\rule[-0.175pt]{0.723pt}{0.350pt}}
\put(578,454){\rule[-0.175pt]{0.723pt}{0.350pt}}
\put(581,455){\rule[-0.175pt]{0.723pt}{0.350pt}}
\put(584,456){\rule[-0.175pt]{0.662pt}{0.350pt}}
\put(586,457){\rule[-0.175pt]{0.662pt}{0.350pt}}
\put(589,458){\rule[-0.175pt]{0.662pt}{0.350pt}}
\put(592,459){\rule[-0.175pt]{0.662pt}{0.350pt}}
\put(595,460){\rule[-0.175pt]{0.723pt}{0.350pt}}
\put(598,461){\rule[-0.175pt]{0.723pt}{0.350pt}}
\put(601,462){\rule[-0.175pt]{0.723pt}{0.350pt}}
\put(604,463){\rule[-0.175pt]{0.723pt}{0.350pt}}
\put(607,464){\rule[-0.175pt]{0.723pt}{0.350pt}}
\put(610,465){\rule[-0.175pt]{0.723pt}{0.350pt}}
\put(613,466){\rule[-0.175pt]{0.723pt}{0.350pt}}
\put(616,467){\rule[-0.175pt]{0.723pt}{0.350pt}}
\put(619,468){\rule[-0.175pt]{0.723pt}{0.350pt}}
\put(622,469){\rule[-0.175pt]{0.723pt}{0.350pt}}
\put(625,470){\rule[-0.175pt]{0.723pt}{0.350pt}}
\put(628,471){\rule[-0.175pt]{0.723pt}{0.350pt}}
\put(631,472){\rule[-0.175pt]{0.723pt}{0.350pt}}
\put(634,473){\rule[-0.175pt]{0.723pt}{0.350pt}}
\put(637,474){\rule[-0.175pt]{0.723pt}{0.350pt}}
\put(640,475){\rule[-0.175pt]{0.723pt}{0.350pt}}
\put(643,476){\rule[-0.175pt]{0.964pt}{0.350pt}}
\put(647,477){\rule[-0.175pt]{0.964pt}{0.350pt}}
\put(651,478){\rule[-0.175pt]{0.964pt}{0.350pt}}
\put(655,479){\rule[-0.175pt]{0.723pt}{0.350pt}}
\put(658,480){\rule[-0.175pt]{0.723pt}{0.350pt}}
\put(661,481){\rule[-0.175pt]{0.723pt}{0.350pt}}
\put(664,482){\rule[-0.175pt]{0.723pt}{0.350pt}}
\put(667,483){\rule[-0.175pt]{0.662pt}{0.350pt}}
\put(669,484){\rule[-0.175pt]{0.662pt}{0.350pt}}
\put(672,485){\rule[-0.175pt]{0.662pt}{0.350pt}}
\put(675,486){\rule[-0.175pt]{0.662pt}{0.350pt}}
\put(678,487){\rule[-0.175pt]{0.723pt}{0.350pt}}
\put(681,488){\rule[-0.175pt]{0.723pt}{0.350pt}}
\put(684,489){\rule[-0.175pt]{0.723pt}{0.350pt}}
\put(687,490){\rule[-0.175pt]{0.723pt}{0.350pt}}
\put(690,491){\rule[-0.175pt]{0.964pt}{0.350pt}}
\put(694,492){\rule[-0.175pt]{0.964pt}{0.350pt}}
\put(698,493){\rule[-0.175pt]{0.964pt}{0.350pt}}
\put(702,494){\rule[-0.175pt]{0.723pt}{0.350pt}}
\put(705,495){\rule[-0.175pt]{0.723pt}{0.350pt}}
\put(708,496){\rule[-0.175pt]{0.723pt}{0.350pt}}
\put(711,497){\rule[-0.175pt]{0.723pt}{0.350pt}}
\put(714,498){\rule[-0.175pt]{0.723pt}{0.350pt}}
\put(717,499){\rule[-0.175pt]{0.723pt}{0.350pt}}
\put(720,500){\rule[-0.175pt]{0.723pt}{0.350pt}}
\put(723,501){\rule[-0.175pt]{0.723pt}{0.350pt}}
\put(726,502){\rule[-0.175pt]{0.964pt}{0.350pt}}
\put(730,503){\rule[-0.175pt]{0.964pt}{0.350pt}}
\put(734,504){\rule[-0.175pt]{0.964pt}{0.350pt}}
\put(738,505){\rule[-0.175pt]{0.662pt}{0.350pt}}
\put(740,506){\rule[-0.175pt]{0.662pt}{0.350pt}}
\put(743,507){\rule[-0.175pt]{0.662pt}{0.350pt}}
\put(746,508){\rule[-0.175pt]{0.662pt}{0.350pt}}
\put(749,509){\rule[-0.175pt]{0.723pt}{0.350pt}}
\put(752,510){\rule[-0.175pt]{0.723pt}{0.350pt}}
\put(755,511){\rule[-0.175pt]{0.723pt}{0.350pt}}
\put(758,512){\rule[-0.175pt]{0.723pt}{0.350pt}}
\put(761,513){\rule[-0.175pt]{0.964pt}{0.350pt}}
\put(765,514){\rule[-0.175pt]{0.964pt}{0.350pt}}
\put(769,515){\rule[-0.175pt]{0.964pt}{0.350pt}}
\put(773,516){\rule[-0.175pt]{0.723pt}{0.350pt}}
\put(776,517){\rule[-0.175pt]{0.723pt}{0.350pt}}
\put(779,518){\rule[-0.175pt]{0.723pt}{0.350pt}}
\put(782,519){\rule[-0.175pt]{0.723pt}{0.350pt}}
\put(785,520){\rule[-0.175pt]{0.723pt}{0.350pt}}
\put(788,521){\rule[-0.175pt]{0.723pt}{0.350pt}}
\put(791,522){\rule[-0.175pt]{0.723pt}{0.350pt}}
\put(794,523){\rule[-0.175pt]{0.723pt}{0.350pt}}
\put(797,524){\rule[-0.175pt]{0.964pt}{0.350pt}}
\put(801,525){\rule[-0.175pt]{0.964pt}{0.350pt}}
\put(805,526){\rule[-0.175pt]{0.964pt}{0.350pt}}
\put(809,527){\rule[-0.175pt]{0.662pt}{0.350pt}}
\put(811,528){\rule[-0.175pt]{0.662pt}{0.350pt}}
\put(814,529){\rule[-0.175pt]{0.662pt}{0.350pt}}
\put(817,530){\rule[-0.175pt]{0.662pt}{0.350pt}}
\put(820,531){\rule[-0.175pt]{0.964pt}{0.350pt}}
\put(824,532){\rule[-0.175pt]{0.964pt}{0.350pt}}
\put(828,533){\rule[-0.175pt]{0.964pt}{0.350pt}}
\put(832,534){\rule[-0.175pt]{0.723pt}{0.350pt}}
\put(835,535){\rule[-0.175pt]{0.723pt}{0.350pt}}
\put(838,536){\rule[-0.175pt]{0.723pt}{0.350pt}}
\put(841,537){\rule[-0.175pt]{0.723pt}{0.350pt}}
\put(844,538){\rule[-0.175pt]{0.723pt}{0.350pt}}
\put(847,539){\rule[-0.175pt]{0.723pt}{0.350pt}}
\put(850,540){\rule[-0.175pt]{0.723pt}{0.350pt}}
\put(853,541){\rule[-0.175pt]{0.723pt}{0.350pt}}
\put(856,542){\rule[-0.175pt]{0.964pt}{0.350pt}}
\put(860,543){\rule[-0.175pt]{0.964pt}{0.350pt}}
\put(864,544){\rule[-0.175pt]{0.964pt}{0.350pt}}
\put(868,545){\rule[-0.175pt]{0.723pt}{0.350pt}}
\put(871,546){\rule[-0.175pt]{0.723pt}{0.350pt}}
\put(874,547){\rule[-0.175pt]{0.723pt}{0.350pt}}
\put(877,548){\rule[-0.175pt]{0.723pt}{0.350pt}}
\put(880,549){\rule[-0.175pt]{0.662pt}{0.350pt}}
\put(882,550){\rule[-0.175pt]{0.662pt}{0.350pt}}
\put(885,551){\rule[-0.175pt]{0.662pt}{0.350pt}}
\put(888,552){\rule[-0.175pt]{0.662pt}{0.350pt}}
\put(891,553){\rule[-0.175pt]{0.964pt}{0.350pt}}
\put(895,554){\rule[-0.175pt]{0.964pt}{0.350pt}}
\put(899,555){\rule[-0.175pt]{0.964pt}{0.350pt}}
\put(903,556){\rule[-0.175pt]{0.723pt}{0.350pt}}
\put(906,557){\rule[-0.175pt]{0.723pt}{0.350pt}}
\put(909,558){\rule[-0.175pt]{0.723pt}{0.350pt}}
\put(912,559){\rule[-0.175pt]{0.723pt}{0.350pt}}
\put(915,560){\rule[-0.175pt]{0.723pt}{0.350pt}}
\put(918,561){\rule[-0.175pt]{0.723pt}{0.350pt}}
\put(921,562){\rule[-0.175pt]{0.723pt}{0.350pt}}
\put(924,563){\rule[-0.175pt]{0.723pt}{0.350pt}}
\put(927,564){\rule[-0.175pt]{0.964pt}{0.350pt}}
\put(931,565){\rule[-0.175pt]{0.964pt}{0.350pt}}
\put(935,566){\rule[-0.175pt]{0.964pt}{0.350pt}}
\put(939,567){\rule[-0.175pt]{0.723pt}{0.350pt}}
\put(942,568){\rule[-0.175pt]{0.723pt}{0.350pt}}
\put(945,569){\rule[-0.175pt]{0.723pt}{0.350pt}}
\put(948,570){\rule[-0.175pt]{0.723pt}{0.350pt}}
\put(951,571){\rule[-0.175pt]{0.662pt}{0.350pt}}
\put(953,572){\rule[-0.175pt]{0.662pt}{0.350pt}}
\put(956,573){\rule[-0.175pt]{0.662pt}{0.350pt}}
\put(959,574){\rule[-0.175pt]{0.662pt}{0.350pt}}
\put(962,575){\rule[-0.175pt]{0.964pt}{0.350pt}}
\put(966,576){\rule[-0.175pt]{0.964pt}{0.350pt}}
\put(970,577){\rule[-0.175pt]{0.964pt}{0.350pt}}
\put(974,578){\rule[-0.175pt]{0.723pt}{0.350pt}}
\put(977,579){\rule[-0.175pt]{0.723pt}{0.350pt}}
\put(980,580){\rule[-0.175pt]{0.723pt}{0.350pt}}
\put(983,581){\rule[-0.175pt]{0.723pt}{0.350pt}}
\put(986,582){\rule[-0.175pt]{0.723pt}{0.350pt}}
\put(989,583){\rule[-0.175pt]{0.723pt}{0.350pt}}
\put(992,584){\rule[-0.175pt]{0.723pt}{0.350pt}}
\put(995,585){\rule[-0.175pt]{0.723pt}{0.350pt}}
\put(998,586){\rule[-0.175pt]{0.723pt}{0.350pt}}
\put(1001,587){\rule[-0.175pt]{0.723pt}{0.350pt}}
\put(1004,588){\rule[-0.175pt]{0.723pt}{0.350pt}}
\put(1007,589){\rule[-0.175pt]{0.723pt}{0.350pt}}
\put(1010,590){\rule[-0.175pt]{0.964pt}{0.350pt}}
\put(1014,591){\rule[-0.175pt]{0.964pt}{0.350pt}}
\put(1018,592){\rule[-0.175pt]{0.964pt}{0.350pt}}
\put(1022,593){\rule[-0.175pt]{0.662pt}{0.350pt}}
\put(1024,594){\rule[-0.175pt]{0.662pt}{0.350pt}}
\put(1027,595){\rule[-0.175pt]{0.662pt}{0.350pt}}
\put(1030,596){\rule[-0.175pt]{0.662pt}{0.350pt}}
\put(1033,597){\rule[-0.175pt]{0.723pt}{0.350pt}}
\put(1036,598){\rule[-0.175pt]{0.723pt}{0.350pt}}
\put(1039,599){\rule[-0.175pt]{0.723pt}{0.350pt}}
\put(1042,600){\rule[-0.175pt]{0.723pt}{0.350pt}}
\put(1045,601){\rule[-0.175pt]{0.723pt}{0.350pt}}
\put(1048,602){\rule[-0.175pt]{0.723pt}{0.350pt}}
\put(1051,603){\rule[-0.175pt]{0.723pt}{0.350pt}}
\put(1054,604){\rule[-0.175pt]{0.723pt}{0.350pt}}
\put(1057,605){\rule[-0.175pt]{0.723pt}{0.350pt}}
\put(1060,606){\rule[-0.175pt]{0.723pt}{0.350pt}}
\put(1063,607){\rule[-0.175pt]{0.723pt}{0.350pt}}
\put(1066,608){\rule[-0.175pt]{0.723pt}{0.350pt}}
\put(1069,609){\rule[-0.175pt]{0.964pt}{0.350pt}}
\put(1073,610){\rule[-0.175pt]{0.964pt}{0.350pt}}
\put(1077,611){\rule[-0.175pt]{0.964pt}{0.350pt}}
\put(1081,612){\rule[-0.175pt]{0.723pt}{0.350pt}}
\put(1084,613){\rule[-0.175pt]{0.723pt}{0.350pt}}
\put(1087,614){\rule[-0.175pt]{0.723pt}{0.350pt}}
\put(1090,615){\rule[-0.175pt]{0.723pt}{0.350pt}}
\put(1093,616){\rule[-0.175pt]{0.723pt}{0.350pt}}
\put(1096,617){\rule[-0.175pt]{0.723pt}{0.350pt}}
\put(1099,618){\rule[-0.175pt]{0.723pt}{0.350pt}}
\put(1102,619){\rule[-0.175pt]{0.723pt}{0.350pt}}
\put(1105,620){\rule[-0.175pt]{0.662pt}{0.350pt}}
\put(1107,621){\rule[-0.175pt]{0.662pt}{0.350pt}}
\put(1110,622){\rule[-0.175pt]{0.662pt}{0.350pt}}
\put(1113,623){\rule[-0.175pt]{0.662pt}{0.350pt}}
\put(1116,624){\rule[-0.175pt]{0.723pt}{0.350pt}}
\put(1119,625){\rule[-0.175pt]{0.723pt}{0.350pt}}
\put(1122,626){\rule[-0.175pt]{0.723pt}{0.350pt}}
\put(1125,627){\rule[-0.175pt]{0.723pt}{0.350pt}}
\put(1128,628){\rule[-0.175pt]{0.723pt}{0.350pt}}
\put(1131,629){\rule[-0.175pt]{0.723pt}{0.350pt}}
\put(1134,630){\rule[-0.175pt]{0.723pt}{0.350pt}}
\put(1137,631){\rule[-0.175pt]{0.723pt}{0.350pt}}
\put(1140,632){\rule[-0.175pt]{0.723pt}{0.350pt}}
\put(1143,633){\rule[-0.175pt]{0.723pt}{0.350pt}}
\put(1146,634){\rule[-0.175pt]{0.723pt}{0.350pt}}
\put(1149,635){\rule[-0.175pt]{0.723pt}{0.350pt}}
\put(1152,636){\rule[-0.175pt]{0.723pt}{0.350pt}}
\put(1155,637){\rule[-0.175pt]{0.723pt}{0.350pt}}
\put(1158,638){\rule[-0.175pt]{0.723pt}{0.350pt}}
\put(1161,639){\rule[-0.175pt]{0.723pt}{0.350pt}}
\put(1164,640){\rule[-0.175pt]{0.723pt}{0.350pt}}
\put(1167,641){\rule[-0.175pt]{0.723pt}{0.350pt}}
\put(1170,642){\rule[-0.175pt]{0.723pt}{0.350pt}}
\put(1173,643){\rule[-0.175pt]{0.723pt}{0.350pt}}
\put(1176,644){\rule[-0.175pt]{0.662pt}{0.350pt}}
\put(1178,645){\rule[-0.175pt]{0.662pt}{0.350pt}}
\put(1181,646){\rule[-0.175pt]{0.662pt}{0.350pt}}
\put(1184,647){\rule[-0.175pt]{0.662pt}{0.350pt}}
\put(1187,648){\rule[-0.175pt]{0.723pt}{0.350pt}}
\put(1190,649){\rule[-0.175pt]{0.723pt}{0.350pt}}
\put(1193,650){\rule[-0.175pt]{0.723pt}{0.350pt}}
\put(1196,651){\rule[-0.175pt]{0.723pt}{0.350pt}}
\put(1199,652){\rule[-0.175pt]{0.723pt}{0.350pt}}
\put(1202,653){\rule[-0.175pt]{0.723pt}{0.350pt}}
\put(1205,654){\rule[-0.175pt]{0.723pt}{0.350pt}}
\put(1208,655){\rule[-0.175pt]{0.723pt}{0.350pt}}
\put(1211,656){\rule[-0.175pt]{0.723pt}{0.350pt}}
\put(1214,657){\rule[-0.175pt]{0.723pt}{0.350pt}}
\put(1217,658){\rule[-0.175pt]{0.723pt}{0.350pt}}
\put(1220,659){\rule[-0.175pt]{0.723pt}{0.350pt}}
\put(1223,660){\rule[-0.175pt]{0.723pt}{0.350pt}}
\put(1226,661){\rule[-0.175pt]{0.723pt}{0.350pt}}
\put(1229,662){\rule[-0.175pt]{0.723pt}{0.350pt}}
\put(1232,663){\rule[-0.175pt]{0.723pt}{0.350pt}}
\put(1235,664){\rule[-0.175pt]{0.723pt}{0.350pt}}
\put(1238,665){\rule[-0.175pt]{0.723pt}{0.350pt}}
\put(1241,666){\rule[-0.175pt]{0.723pt}{0.350pt}}
\put(1244,667){\rule[-0.175pt]{0.723pt}{0.350pt}}
\put(1247,668){\rule[-0.175pt]{0.662pt}{0.350pt}}
\put(1249,669){\rule[-0.175pt]{0.662pt}{0.350pt}}
\put(1252,670){\rule[-0.175pt]{0.662pt}{0.350pt}}
\put(1255,671){\rule[-0.175pt]{0.662pt}{0.350pt}}
\put(1258,672){\rule[-0.175pt]{0.723pt}{0.350pt}}
\put(1261,673){\rule[-0.175pt]{0.723pt}{0.350pt}}
\put(1264,674){\rule[-0.175pt]{0.723pt}{0.350pt}}
\put(1267,675){\rule[-0.175pt]{0.723pt}{0.350pt}}
\put(1270,676){\rule[-0.175pt]{0.723pt}{0.350pt}}
\put(1273,677){\rule[-0.175pt]{0.723pt}{0.350pt}}
\put(1276,678){\rule[-0.175pt]{0.723pt}{0.350pt}}
\put(1279,679){\rule[-0.175pt]{0.723pt}{0.350pt}}
\put(1282,680){\rule[-0.175pt]{0.578pt}{0.350pt}}
\put(1284,681){\rule[-0.175pt]{0.578pt}{0.350pt}}
\put(1286,682){\rule[-0.175pt]{0.578pt}{0.350pt}}
\put(1289,683){\rule[-0.175pt]{0.578pt}{0.350pt}}
\put(1291,684){\rule[-0.175pt]{0.578pt}{0.350pt}}
\put(1294,685){\rule[-0.175pt]{0.723pt}{0.350pt}}
\put(1297,686){\rule[-0.175pt]{0.723pt}{0.350pt}}
\put(1300,687){\rule[-0.175pt]{0.723pt}{0.350pt}}
\put(1303,688){\rule[-0.175pt]{0.723pt}{0.350pt}}
\put(1306,689){\rule[-0.175pt]{0.723pt}{0.350pt}}
\put(1309,690){\rule[-0.175pt]{0.723pt}{0.350pt}}
\put(1312,691){\rule[-0.175pt]{0.723pt}{0.350pt}}
\put(1315,692){\rule[-0.175pt]{0.723pt}{0.350pt}}
\put(1318,693){\rule[-0.175pt]{0.662pt}{0.350pt}}
\put(1320,694){\rule[-0.175pt]{0.662pt}{0.350pt}}
\put(1323,695){\rule[-0.175pt]{0.662pt}{0.350pt}}
\put(1326,696){\rule[-0.175pt]{0.662pt}{0.350pt}}
\put(1329,697){\rule[-0.175pt]{0.578pt}{0.350pt}}
\put(1331,698){\rule[-0.175pt]{0.578pt}{0.350pt}}
\put(1333,699){\rule[-0.175pt]{0.578pt}{0.350pt}}
\put(1336,700){\rule[-0.175pt]{0.578pt}{0.350pt}}
\put(1338,701){\rule[-0.175pt]{0.578pt}{0.350pt}}
\put(1341,702){\rule[-0.175pt]{0.723pt}{0.350pt}}
\put(1344,703){\rule[-0.175pt]{0.723pt}{0.350pt}}
\put(1347,704){\rule[-0.175pt]{0.723pt}{0.350pt}}
\put(1350,705){\rule[-0.175pt]{0.723pt}{0.350pt}}
\put(1353,706){\rule[-0.175pt]{0.723pt}{0.350pt}}
\put(1356,707){\rule[-0.175pt]{0.723pt}{0.350pt}}
\put(1359,708){\rule[-0.175pt]{0.723pt}{0.350pt}}
\put(1362,709){\rule[-0.175pt]{0.723pt}{0.350pt}}
\put(1365,710){\rule[-0.175pt]{0.578pt}{0.350pt}}
\put(1367,711){\rule[-0.175pt]{0.578pt}{0.350pt}}
\put(1369,712){\rule[-0.175pt]{0.578pt}{0.350pt}}
\put(1372,713){\rule[-0.175pt]{0.578pt}{0.350pt}}
\put(1374,714){\rule[-0.175pt]{0.578pt}{0.350pt}}
\put(1377,715){\rule[-0.175pt]{0.723pt}{0.350pt}}
\put(1380,716){\rule[-0.175pt]{0.723pt}{0.350pt}}
\put(1383,717){\rule[-0.175pt]{0.723pt}{0.350pt}}
\put(1386,718){\rule[-0.175pt]{0.723pt}{0.350pt}}
\put(1389,719){\rule[-0.175pt]{0.530pt}{0.350pt}}
\put(1391,720){\rule[-0.175pt]{0.530pt}{0.350pt}}
\put(1393,721){\rule[-0.175pt]{0.530pt}{0.350pt}}
\put(1395,722){\rule[-0.175pt]{0.530pt}{0.350pt}}
\put(1397,723){\rule[-0.175pt]{0.530pt}{0.350pt}}
\put(1399,724){\rule[-0.175pt]{0.723pt}{0.350pt}}
\put(1403,725){\rule[-0.175pt]{0.723pt}{0.350pt}}
\put(1406,726){\rule[-0.175pt]{0.723pt}{0.350pt}}
\put(1409,727){\rule[-0.175pt]{0.723pt}{0.350pt}}
\put(1412,728){\rule[-0.175pt]{0.578pt}{0.350pt}}
\put(1414,729){\rule[-0.175pt]{0.578pt}{0.350pt}}
\put(1416,730){\rule[-0.175pt]{0.578pt}{0.350pt}}
\put(1419,731){\rule[-0.175pt]{0.578pt}{0.350pt}}
\put(1421,732){\rule[-0.175pt]{0.578pt}{0.350pt}}
\put(1424,733){\rule[-0.175pt]{0.723pt}{0.350pt}}
\put(1427,734){\rule[-0.175pt]{0.723pt}{0.350pt}}
\put(1430,735){\rule[-0.175pt]{0.723pt}{0.350pt}}
\put(1433,736){\rule[-0.175pt]{0.723pt}{0.350pt}}
\sbox{\plotpoint}{\rule[-0.250pt]{0.500pt}{0.500pt}}%
\put(1389,572){\makebox(0,0){$+$}}
\put(824,475){\makebox(0,0){$+$}}
\put(543,416){\makebox(0,0){$+$}}
\put(421,373){\makebox(0,0){$+$}}
\sbox{\plotpoint}{\rule[-0.350pt]{0.700pt}{0.700pt}}%
\put(623,447){\raisebox{-1.2pt}{\makebox(0,0){$\Box$}}}
\put(772,502){\raisebox{-1.2pt}{\makebox(0,0){$\Box$}}}
\put(894,547){\raisebox{-1.2pt}{\makebox(0,0){$\Box$}}}
\put(1017,590){\raisebox{-1.2pt}{\makebox(0,0){$\Box$}}}
\put(1131,605){\raisebox{-1.2pt}{\makebox(0,0){$\Box$}}}
\put(1180,659){\raisebox{-1.2pt}{\makebox(0,0){$\Box$}}}
\put(1270,686){\raisebox{-1.2pt}{\makebox(0,0){$\Box$}}}
\put(1395,720){\raisebox{-1.2pt}{\makebox(0,0){$\Box$}}}
\sbox{\plotpoint}{\rule[-0.175pt]{0.350pt}{0.350pt}}%
\put(623,447){\makebox(0,0){$\times$}}
\put(772,502){\makebox(0,0){$\times$}}
\put(894,547){\makebox(0,0){$\times$}}
\put(1017,590){\makebox(0,0){$\times$}}
\put(1131,605){\makebox(0,0){$\times$}}
\put(1180,659){\makebox(0,0){$\times$}}
\put(1270,686){\makebox(0,0){$\times$}}
\put(1395,720){\makebox(0,0){$\times$}}
\put(623,438){\rule[-0.175pt]{0.350pt}{4.095pt}}
\put(613,438){\rule[-0.175pt]{4.818pt}{0.350pt}}
\put(613,455){\rule[-0.175pt]{4.818pt}{0.350pt}}
\put(772,495){\rule[-0.175pt]{0.350pt}{3.373pt}}
\put(762,495){\rule[-0.175pt]{4.818pt}{0.350pt}}
\put(762,509){\rule[-0.175pt]{4.818pt}{0.350pt}}
\put(894,529){\rule[-0.175pt]{0.350pt}{8.672pt}}
\put(884,529){\rule[-0.175pt]{4.818pt}{0.350pt}}
\put(884,565){\rule[-0.175pt]{4.818pt}{0.350pt}}
\put(1017,586){\rule[-0.175pt]{0.350pt}{1.686pt}}
\put(1007,586){\rule[-0.175pt]{4.818pt}{0.350pt}}
\put(1007,593){\rule[-0.175pt]{4.818pt}{0.350pt}}
\put(1131,587){\rule[-0.175pt]{0.350pt}{8.672pt}}
\put(1121,587){\rule[-0.175pt]{4.818pt}{0.350pt}}
\put(1121,623){\rule[-0.175pt]{4.818pt}{0.350pt}}
\put(1180,644){\rule[-0.175pt]{0.350pt}{7.227pt}}
\put(1170,644){\rule[-0.175pt]{4.818pt}{0.350pt}}
\put(1170,674){\rule[-0.175pt]{4.818pt}{0.350pt}}
\put(1270,677){\rule[-0.175pt]{0.350pt}{4.336pt}}
\put(1260,677){\rule[-0.175pt]{4.818pt}{0.350pt}}
\put(1260,695){\rule[-0.175pt]{4.818pt}{0.350pt}}
\put(1395,703){\rule[-0.175pt]{0.350pt}{7.950pt}}
\put(1385,703){\rule[-0.175pt]{4.818pt}{0.350pt}}
\put(1385,736){\rule[-0.175pt]{4.818pt}{0.350pt}}
\put(1336,566){\usebox{\plotpoint}}
\put(1336,566){\rule[-0.175pt]{0.350pt}{2.891pt}}
\put(1336,572){\usebox{\plotpoint}}
\put(1336,572){\rule[-0.175pt]{24.090pt}{0.350pt}}
\put(780,469){\usebox{\plotpoint}}
\put(780,469){\rule[-0.175pt]{0.350pt}{2.891pt}}
\put(780,475){\usebox{\plotpoint}}
\put(780,475){\rule[-0.175pt]{21.199pt}{0.350pt}}
\put(868,469){\usebox{\plotpoint}}
\put(868,469){\rule[-0.175pt]{0.350pt}{2.891pt}}
\put(513,410){\usebox{\plotpoint}}
\put(513,410){\rule[-0.175pt]{0.350pt}{2.650pt}}
\put(513,416){\usebox{\plotpoint}}
\put(513,416){\rule[-0.175pt]{14.213pt}{0.350pt}}
\put(572,410){\usebox{\plotpoint}}
\put(572,410){\rule[-0.175pt]{0.350pt}{2.650pt}}
\put(403,368){\usebox{\plotpoint}}
\put(403,368){\rule[-0.175pt]{0.350pt}{2.650pt}}
\put(403,373){\usebox{\plotpoint}}
\put(403,373){\rule[-0.175pt]{8.672pt}{0.350pt}}
\put(439,368){\usebox{\plotpoint}}
\put(439,368){\rule[-0.175pt]{0.350pt}{2.650pt}}
\end{picture}

\caption{The running coupling versus a length scale}
\label{fig:dorunrun}
\end{center}
\end{figure*}
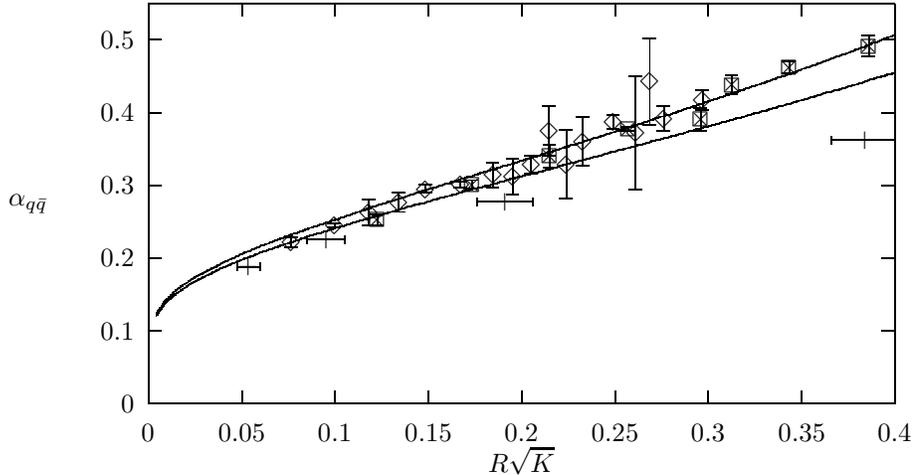

We therefore compare these results with those obtained
previously~\cite{beta27} on $32^4$ lattices at $\beta=2.7$, plotting
$\alpha_{q\bar q}$ against $R$ and rescaling only the horizontal axis.  That
the agreement is entirely comfortable, with no indication of scaling
violation, will become clear when we discuss figure~\ref{fig:dorunrun}
which imposes the even stronger condition that the rescaling factor is
$\sqrt{K}$.

For now, we show in table~\ref{tab:aratio} various values for the ratios
of the lattice spacings
$a(\beta\mathord{=}2.7)/a(\beta\mathord{=}2.85)$.  The first value is
obtained by a full fit to the Monte Carlo results, including an estimate
of the systematic error.  The last is the naive expectation of 2-loop
perturbation theory using the bare couplings.  It is clear there is some
discrepancy.  The other two values require some comment.

\begin{table}
\caption{The ratio $a(2.7)/a(2.85)$ in various calculations}
\begin{tabular}{cccc}
\hline
\vbox to 0pt{\vskip 0.5\baselineskip\hbox{UKQCD}\vss} &
  $\beta_E$ & L\"uscher & Asymptotic \\
 & scheme & et al.~\cite{lusch} & (two loops) \\
\hline
1.60(6) & 1.53 & 1.48 & 1.46 \\
\hline
\end{tabular}
\label{tab:aratio}
\end{table}

It is useful to use instead of the bare coupling something more closely
related to physical processes. One popular choice is the
$\beta_E$-scheme~\cite{betae}, where the plaquette expectation value
$\langle U_{\rm pl}\rangle$ is used as the basis for calculating the
coupling.  Here we use the simplest variant where $\beta_E =
c_2/(1-\langle U_{\rm pl}\rangle)$, in which $c_2$ is a perturbative
coefficient.  Using even this simple correction changes the $a$-ratio by
about half of the discrepancy between Monte Carlo and na\"\i ve
perturbation theory, giving some confidence our results are in the right
region.

The remaining estimate is taken from the fit in figure~3 of
ref.~\cite{lusch}.  In this paper (see the talk by Martin L\"uscher at
this conference) the authors calculate the running coupling by a
recursive finite-size-scaling technique.  The value shown is in terms of
the {\em bare} coupling with lattice sizes determined at fixed {\em
physical} coupling.  The result quoted, close to the two-loop value, is
not central to ref.~\cite{lusch}; it is indicative of the fact that the
starting point for that paper is in small (and therefore perturbative)
volumes.

We have also compared the new results with those on $24^4$ lattices at
$\beta=2.4$.  Even here we find excellent agreement, showing scaling
over a range of lattice spacings
$a(\beta\mathord{=}2.4)/a(\beta\mathord{=}2.85)=4.12(2)$, which is
certainly impressive.

\section{Running coupling}
Equation~\ref{eq:alpha} defines a running coupling in the $\LR$-scheme;
$\LR=1.048\LMSB$ so these schemes are perturbatively close.  We plot
$\alpha_{q\bar q}$ against the dimensionless $R\sqrt K$ which gives a
physical length scale.  This plot is shown in figure~\ref{fig:dorunrun}.

In the figure we show results for both $\beta = 2.85$ ($\Diamond$) and
$\beta = 2.7$ ($\Box$).  For clarity only the larger of the two error
bars (statistical and systematic) is shown here.  (Both sets appear in
ref.~\cite{us}.) The effect of running is clear and the curve appears to
be universal.

The solid lines are plots of $\alpha_{q\bar q}$ from perturbation
theory:
\begin{eqnarray}
\lefteqn{\alpha_{q\bar q}(R)=} \nonumber \\
\lefteqn{{1\over 4\pi\left(b_0\ln(\LR R)^{-2}
 +(b_1/b_0)\ln\ln(\LR R)^{-2}\right)}}
\end{eqnarray}
with the usual perturbative coefficients.  The upper line corresponds to
$a\LR=0.044$ and the lower to $a\LR=0.038$; note that our results
include increasing contributions from the linear potential as $R$
increases which will tend to make our results larger than the
perturbative estimate. From these limits our estimate of the scale
parameter is $a\LR=0.041(3)$.  In terms of the string tension this is
$\sqrt{K}/\LR=1.54(15)$ or $\sqrt{K}/\LMSB=1.62(14)$; using the (in this
case entirely unjustifiable) value $\sqrt{K} = 0.44\,{\rm GeV}$ we get
$\LMSB=272(24)\,{\rm MeV}$.  (In these units our shortest-distance
point, at $R=\sqrt{2}a$, would correspond to a momentum scale
$q=1/R\approx4.4\,{\rm GeV}$.)

The remaining points with horizontal error bars are converted from
ref.~\cite{lusch}.  We have used the value of the string tension
calculated here to set the scale, however the error in the string
tension is not included.  (This is consistent with the other points in
the figure.)  The region of interest is at the lower end; the error bars
just fail to overlap our lines for $\LR$, so the conclusions are
not entirely clear.  If our estimate of systematic errors is to be
believed there appears to be some discrepancy.

\section{Concluding remarks}
It has recently become clear~\cite{lepmac} that attempts to compare
Monte Carlo results from the lattice with perturbation theory using the
bare coupling are doomed.  Here we have presented confirmatory evidence
which further suggests that when one uses more physical quantities, the
usual Wilson theory for SU(2) without fermions is able to produce
results entirely in accord with those in the continuum.   Our bare
coupling is not able to give us agreement with perturbation theory, but
provided we fit to the $\Lambda$-parameter ourselves we see a running
coupling entirely in agreement with the asymptotic two-loop result.  In
addition we see scaling of $\pot$ results between lattices differing by
a factor of four in lattice spacing.

In ref.~\cite{fhk} these new results are displayed in the $\beta_E$
scheme referred to above and it is clearly seen that the approach to
asymptotic behaviour is much smoother when the bare coupling is not
used, enabling an extrapolated value of the ratio $\sqrt{K}/\LMSB$ to be
extracted. Their value is $1.79(12)$, slightly higher than our
unextrapolated number which is $1.62(14)$.

Overall, we believe that pure-gauge lattice SU(2) is already able to
give results which accurately reflect continuum dynamics.

\end{document}